\documentstyle[12pt]{article}
\newcommand{\fit}{\hat{F}^i(\th)}
\newcommand{\beq}{\begin{equation}}
\newcommand{\eeq}{\end{equation}}
\newcommand{\bdm}{\begin{displaymath}}
\newcommand{\edm}{\end{displaymath}}
\newcommand{\bea}{\begin{eqnarray}}
\newcommand{\eea}{\end{eqnarray}}

\newcommand{\th}{\theta}

\newcommand{\kb}{\bar{k}}

\newcommand{\ib}{\bar{\imath}}

\begin{document}
\setcounter{page}{0}
\topmargin 0pt
\oddsidemargin 5mm
\renewcommand{\thefootnote}{\fnsymbol{footnote}}
\newpage
\setcounter{page}{0}
\begin{titlepage}
\begin{flushright}
Swansea SWAT/93-94/30 \\
hep-th/9405034
\end{flushright}
\vspace{0.5cm}
\begin{center}
{\Large {\bf Vertex Operators and Soliton Time Delays in Affine Toda Field
Theory}} \\
\vspace{1.8cm}
{\large A. Fring, P.R. Johnson, M.A.C. Kneipp and D.I. Olive }

{\em Departments of Physics and Mathematics,\\
University College of Swansea,
Swansea SA2 8PP,UK.} \\
\vspace{2cm}
\renewcommand{\thefootnote}{\arabic{footnote}}
\setcounter{footnote}{0}
\begin{abstract}
In a space-time of two dimensions the overall effect of the collision of two
solitons is a time delay (or advance) of their final trajectories relative to
their
initial trajectories. For the solitons of affine Toda field theories, the
space-time
displacement of the trajectories is proportional to the logarithm of a number
$X$ depending only on the species of the colliding solitons and their rapidity
difference. $X$ is the factor arising in the normal ordering of the product of
the
two vertex operators associated with the solitons. $X$ is shown to take real
values between $0$ and $1$. This means that, whenever the solitons are
distinguishable, so that transmission rather than reflection is the only
possible interpretation of the classical scattering process, the time delay is
negative and so an indication of attractive forces between the solitons.
\end{abstract}
\vspace{.3cm}
\centerline{April 1994}
 \end{center}
\end{titlepage}
\newpage
%******************************************************************
\section{Introduction}

Affine Toda field theories \cite{MOP} are relativistically invariant field
theories which
are integrable in a space-time of two dimensions and possess a natural
interpretation as special deformations of conformally invariant theories
\cite{EY,Z,HM}.
When the coupling is imaginary so that there are degenerate vacua, the
equations support solutions describing any number of solitons interpolating
the vacua. A number of authors have worked out examples on a case by case
basis \cite{Hol2,MM,ACFGZ,CZ}. On the other hand, a general formalism for these
solutions has recently been found
\cite{OTUa,OTUb} exploiting a basis of the underlying affine Kac-Moody algebra
in which the
principal Heisenberg subalgebra plays a significant r\^{o}le. This subalgebra
is isomorphic to the algebra of conserved charges or ``energies'' and can
be thought of as an infinite Poincar\'{e} algebra appropriate to an integrable
theory. The simplest such theory, namely that associated with affine $su(2)$,
is very familiar as sine-Gordon theory \cite{FK,Raj}.

In the formalism, the individual solitons are ``created'' by group elements
obtained by exponentiating quantities $\hat{F}^1, \hat{F}^2,\ldots, \hat{F}^r$
which ad-diagonalise the  ``energies'' generating the Heisenberg subalgebra.
Each exponential series terminates with the highest non-vanishing power of
$\hat{F}^i$ being expressible as a vertex operator obtained by exponentiating
and normal ordering an element of the Heisenberg subalgebra \cite{OTUb,KOa}
when the affine Kac-Moody algebra is untwisted and simply laced (and also
when it is twisted \cite{KOb}).
This result is sufficient to show that these solutions correctly interpolate
degenerate vacua.

In this paper we show that these vertex operators determine yet more
detail of the asymptotic behaviour of the soliton solutions. In these
solutions the energy-momentum vector of a specific soliton is unchanged by
collision but the trajectory may sustain a lateral displacement in space-time
as discussed in section 2. Traditionally this is parametrised by the
time delay in the centre of momentum frame. After a review of the vertex
operator
formalism in section 3, our first result, in section 4, is that this lateral
displacement can be expressed straightforwardly in terms of the logarithms of
the numbers $X_{ik}(\theta_i-\theta_k)$ arising in the procedure of normal
ordering the product of two vertex operators mentioned above as being
associated
with
solitons $i$ and $k$.

In section 5 the overall lateral displacement of the soliton trajectories
due to the scattering with several other solitons is determined and shown
to be independent of the temporal order in which the collisions take place.
This is because the displacement is simply additive. The result constitutes
the classical analogue of the Yang-Baxter equation \cite{YB} and bootstrap
equations \cite{ZZ} for the quantum scattering matrix featuring the
factorisation property of the $n$ particle S-matrix into two particle
S-matrices.

In section 6 various properties of the number $X_{ik}(\theta_i-\theta_k)$
are established, including symmetry and crossing properties. In particular
it is verified that it is real when the rapidity difference is real, as the
physical interpretation demands. Furthermore it is shown to take values
restricted to lie between $0$ and $1$, so that the associated time delay
(in the centre of momentum frame) is always negative. Suppose two
distinguishable solitons are considered in the sense that they carry different
species or different topological charges. In this case the solution
describing the scattering has to be regarded as a transmission rather than a
reflection. If the solitons are indistinguishable either interpretation
is possible. With this understanding our result indicates that the forces
between two distinguishable solitons are always attractive because of the
time advance.

In the concluding section 7, we mention the well known connection between
the time delay and the semi-classical approximation to the S-matrix as well
as the intriguing similarity of the structure of $X_{ik}(\theta_i-\theta_k)$
with the known scattering matrices in affine Toda field theories
\cite{Dor2,FO}.

\section{Kinematics of scattering of two particles in two dimensions}
\setcounter{equation}{0}
In an integrable theory in two dimensions, when two particles collide
the outcome consists of two particles with the same masses as the
original particles. If the two masses differ, the corresponding
energy-momentum vectors are unchanged. If the two masses are equal, even
though the particles are distinct, it is
kinematically possible for the energy-momentum vectors to interchange.

Classically the particles describe trajectories in space-time which are
straight except near the collision. The collision may displace the trajectories
laterally but as the energy-momentum is unchanged, the final direction
coincides with
the initial direction. We now consider alternative descriptions of the
displacements of the trajectories and show how the conservation laws correlate
the displacement of the trajectories of the two colliding particles.

Consider first a single particle with velocity $v$, energy $E$, and hence
momentum $vE$. Before collision the equation of the trajectory in space-time
is
\beq
x=vt+x(I), \label{eq: 2.1}\eeq
whereas afterwards it is
\beq
x=vt+x(F), \label{eq: 2.2}\eeq
as the velocity is unchanged. Only the intercept with the $x$-axis changes. So
\beq
\Delta(x)=x(F)-x(I)  \label{eq: 2.3} \eeq
measures the lateral displacement at fixed time. This is not Lorentz invariant
but the combination $E\Delta(x)$ is. Since $E$ is always positive  it follows
that $\Delta(x)$ has the same sign in all Lorentz frames of reference, even
though its magnitude varies.
The intercepts of the trajectories (\ref{eq: 2.1}) and  (\ref{eq: 2.2}) with
the time axis are given by
\beq
t(I)=-{x(I)\over v},\qquad t(F)=-{x(F)\over v}. \eeq
Then, we define the ``time delay''
\beq
\Delta(t)=t(F)-t(I)=-{\Delta(x)\over v}. \label{eq: 2.5} \eeq
Again
\beq
E\Delta(x)=-p\Delta(t) \label{eq: 2.6} \eeq
is Lorentz invariant. As the sign of $p$ can be changed by a Lorentz
transformation so can that of $\Delta(t)$.

Now consider both particles participating in the collision, labelling them
$1$ and $2$. Consider the ``centre of energy'' coordinate
$$
X={E_{1}x_{1}+E_{2}x_{2} \over E_{1}+E_{2}}. $$
Then
$${dX\over dt}={p_{1}+p_{2}\over E_{1}+E_{2}}$$
is constant throughout time so that
$$ X={p_{1}+p_{2}\over E_{1}+E_{2}}t + X_{0}. $$
Now compare the results of inserting the trajectories (\ref{eq: 2.1}) before
the collision with the result of inserting (\ref{eq: 2.2}) after the
collision. As the results must agree
$$
E_{1}\Bigl(x_{1}(F)-x_{1}(I)\Bigr)+E_{2}\Bigl(x_{2}(F)-x_{2}(I)\Bigr)=0$$
or, denoting $\Delta_{12}(x)=x_{1}(F)-x_{1}(I)$ and similarly for
$\Delta_{21}(x)$
\beq
E_{1}\Delta_{12}(x)+E_{2}\Delta_{21}(x)=0. \label{eq: 2.7} \eeq
Thus the spatial displacement of the two trajectories must have opposite signs.
By (\ref{eq: 2.6}) we have, equally,
\beq
p_{1}\Delta_{12}(t)+p_{2}\Delta_{21}(t)=0, \eeq
where $\Delta_{12}(t)=t_{1}(F)-t_{1}(I)$ is the time delay sustained by
particle 1 colliding with particle 2. Notice that, in the centre of momentum
frame $p_{1}+p_{2}=0$, the two time delays are equal:
\beq
\Delta_{12}(t)=\Delta_{21}(t). \eeq

We see from (2.6), (2.7) and (2.8) that if particle 1 moves faster than
particle 2, then the three quantities (2.9), $\Delta_{21}(x)$ and
$-\Delta_{12}(x)$ all have the same sign. This common sign has a physical
interpretation. Suppose the force between the particles is attractive.
Then particle 1 will accelerate as it approaches particle 2 and afterwards
decelerate. As a result $\Delta_{12}(x)$ will be positive and the common
sign negative. Thus an attractive force implies a negative time delay, in
other words a time advance, in the centre of momentum frame.
A repulsive force implies a time delay, if there is transmission. There is
also the additional possibility of a reflection with either a delay or
advance if the two masses are equal.

The preceding discussion of relativistic particles colliding classically
applies also to relativistic solitons and, in particular, to the solitons
of affine Toda field theory. In the subsequent sections we shall show that
when soliton 1 collides with soliton 2 the resultant displacements are given
by
\beq
E_{1}\Delta_{12}(x)=-p_{1}\Delta_{12}(t)=-{\rm sign}(p_{1}-p_{2})
{2h\over |\beta |^2}\ln X_{12}(\theta_1-\theta_2) \label{eq: 2.10} \eeq
where $h$ is the Coxeter number associated with the theory and $|\beta |$ the
magnitude of the imaginary coupling constant $\beta$. The quantity
$X_{12}(\theta_1-\theta_2)$ depending on the rapidity difference
$\th=\theta_1-\theta_2$ of the two solitons has been met before. It occurs when
the product of the two vertex operators associated with the solitons $1$ and
$2$ are normal ordered \cite{OTUb}. Equation (\ref{eq: 2.10}) generalises
the well known result
for the time delay in sine-Gordon theory \cite{PS,Rub} when
$$X_{12}(\th)=\tanh^2\left({\th\over 2}\right).$$

Thus $X_{12}(\theta)$ has acquired a new physical interpretation whose
viability requires it to possess various properties not hitherto apparent.
For example we shall show that $X_{12}(\theta)$ is real when $\theta$ is real
and that it satisfies the symmetry property
$$X_{12}(\theta)=X_{21}(\theta) $$
demanded by (\ref{eq: 2.7}). Furthermore it takes values between $0$ and $1$.
Hence, in the centre of momentum frame, the time delay is always negative by
(\ref{eq: 2.10}) whatever the velocities concerned.
As explained, this suggests that affine Toda solitons exert
attractive forces on each other. The only possible exception to this is when
two identical solitons are considered, with the same species and
topological quantum number. Then it is possible to interpret the scattering
as a reflection rather than a transmission. In this case it is possible for
the force to be repulsive. This is the accepted point of view in sine-Gordon
theory which furnishes a special case of our result (\ref{eq: 2.10}) \cite{JW}.

\section{Soliton solutions and vertex operators}
\setcounter{equation}{0}
Here we shall recall the general formalism for soliton solutions in affine
Toda field theory and the r\^{o}le played by vertex operators, at least when
the
associated affine Kac-Moody algebra is untwisted and simply laced. The
extension to the twisted case is straightforward in view of the work of
\cite{KOb} and to the untwisted non simply laced case only slightly more
complicated.

When the coupling constant $\beta$ is purely imaginary the affine Toda field
theories possess classical solutions describing any number, $N$, say of
solitons which may be composed of any of the $r={\rm rank\ }g$ species
(where $\hat{g}$ is the associated affine Kac-Moody algebra).
The solution takes the form
\beq
e^{-\beta\lambda_{j}.\phi}={\langle\Lambda_{j}|g(t)|\Lambda_{j}\rangle\over
\langle\Lambda_{0}|g(t)|\Lambda_{0}\rangle^{m_j}}. \label{eq: 3.1} \eeq
$\phi(x,t)$ is the $r$ component affine Toda field. $\lambda_{j}$ is the
$j^{\rm th}$ fundamental weight of the finite dimensional Lie algebra $g$,
while $\Lambda_j$ is the corresponding weight of $\hat{g}$, following the
notation of \cite{GO}.
$|\Lambda_j\rangle $ is the highest weight of the
corresponding ``highest weight" representation whose level is
$m_j$. $ \Lambda_0  $ denotes the zero-th fundamental
weight of $\hat g$. $|\Lambda_0\rangle $ can be regarded as
the vacuum state at level $m_0 = 1$. The Kac-Moody group
element $g(t)$ in (\ref{eq: 3.1}) contains the soliton data:
it factorises into $N$ factors, each one characteristic of
each individual soliton
\beq
g(t) = g_N(t) g_{N-1}(t) \cdots g_1(t)
\label{eq: 3.2}
\eeq
where
\beq
g_m(t) = e^{Q_m W_{i(m)}(\th_m) \hat{F}^{i(m)}(\th_m)}
\label{eq: 3.3}
\eeq
are the factors in (\ref{eq: 3.2}) and ordered according to the
rapidities $\th_m$. The real functions
$W_{i(m)}(\th_m)$ carry the dependence on the space, $x$,
and time, $t$, variables:
\beq
 W_{i(m)}(\th_m) = e^{\mu_{i(m)} \left( x \cosh \th_m - t
\sinh
\th_m \right)} .
\label{eq: 3.4}
\eeq
The $m^{ \rm th}$ soliton has ``species" $i(m)$ and rapidity
$\th_m$. When quantised, the affine Toda field $\phi$ creates $r$
species of particles whose masses are $\hbar \mu_1 ,\hbar
\mu_2 , \ldots, \hbar \mu_m$. Thus (\ref{eq: 3.4}) provides
a precise correspondence between the $r$ species of solitons
and the $r$ species of field excitation particle. When $g$
is simply laced so that all roots can be taken to have
length $\sqrt{2}$, the ratios of the masses of corresponding
soliton and field excitation particle are independent of
the species $i$. It has been shown \cite{Hol2,OTUa} that the
mass of the $i$'th species of soliton
\beq
M_i = {2 h \mu_i \over |\beta|^2} .
\label{eq: 3.5}
\eeq
A similar result holds for the twisted theories\cite{KOb}.
The quantities $\hat{F}^i(\th)$ are generators of $\hat{g}$
which ad-diagonalise the principal Heisenberg subalgebra
\beq
\left[\hat{E}_M , \hat{F}^i(\th) \right] = \gamma_i\cdot q([M])
(z_i)^M
\hat{F}^i(\th)
\label{eq: 3.6}
\eeq
in the notation of \cite{OTUa}. The elements of the principal
Heisenberg subalgebra are graded by $d'= T^3 - hL_0$, the
``principal" grade:
\beq
\left[ \hat{E}_M , \hat{E}_N \right] = xM\delta_{M+N,0} \ \ \
\ \ \   , \ \ \ \ \  \left[ d' , \hat{E}_M \right] = M
\hat{E}_M.
\label{eq: 3.7}
\eeq
Here $x$ is the level of the representation considered and
it is understood that the $M$ can only equal an exponent of
$\hat{g}$ that is an exponent of $g$ modulo its Coxeter
number $h$. The complex number $z_{i(m)}$ in (\ref{eq: 3.6}) is
related to the rapidity $\th_m$ by
\beq
z_{i(m)} = i e^{-\th_m} e^{-i\pi {\left(1 + c(i) \right) \over
2h}}
\label{eq: 3.8}
\eeq
where the phase ensures that $W_{i(m)}$ is real. The complex
number $Q_m$ can be parametrised as
\beq
Q_m = e^{i\psi_m} e^{-\mu_{i(m)} x^0_m \cosh \th_m }
\label{eq: 3.9}
\eeq
where $x^0_m $ relates to the space coordinate of the $m'$th
soliton at $t=0$ in a way that will be clarified later. The
phase $\psi_m$ relates to the topological quantum number
defined as the difference between the values the affine Toda
field  takes at large distances. Certain discrete values are
forbidden by the requirement that the solution
(\ref{eq: 3.1}) should not develop singularities as $x$
varies over space.

These soliton solutions exhibit a number of important
features. Despite the imaginary nature of $\beta$, the
energy and momentum of the solution (\ref{eq: 3.1}) has been
evaluated and shown to be real and finite (with positive
energy) \cite{OTUa}. Moreover the resulting form is characteristic of $N$
solitons
moving with the stated rapidities and masses (\ref{eq: 3.5}).
That the affine Toda field interpolates degenerate vacua at
large distances can be confirmed explicitly (when $\hat{g}$ is
simply laced \cite{KOa}) using the generalised vertex
operator construction \cite{OTUb, KOa} which we now explain
in more detail.

Consider the single exponential factor (\ref{eq: 3.3})
creating the $m'$th soliton. Since $\fit$ is a
generator of $\hat{g}$, we must check that the exponential
indeed makes sense as a finite operator in representations
of the highest weight considered in  (\ref{eq: 3.1}).
Expanding the exponential as a series, we find that powers
of $\fit$ higher than the level vanish identically  if
$\hat{g}$ is simply laced  \cite{OTUb}. Furthermore, the
highest non vanishing power, namely the level, is given by a
vertex operator obtained by normal ordering an exponential
expression of the principal Heisenberg subalgebra \cite{KOa}:
\beq
{\left(\fit\right)^{m_j} \over (m_j)!} = e^{-2\pi i
\lambda_i\cdot \lambda_j} Y^i_- Y^i_+
\label{eq: 3.10}
\eeq
where
\beq
Y^i_\pm = \exp\left\{ \sum_{M>0}{ {\gamma_i \cdot q(\mp[M])
z^{\mp M} \over \mp M } \hat{E}_{\pm M}} \right\}  ,
\label{eq: 3.11}
\eeq
and $z$ is related to the rapidity $\th$ as in
(\ref{eq: 3.8}). It follows that
\beq
e^{QW_i(\th) \fit} = 1 + \ldots + \left(QW_i(\th)\right)^{m_j}
e^{-2\pi i
\lambda_i\cdot \lambda_j} Y^i_- Y^i_+ .
\label{eq: 3.12}
\eeq
The coefficient of the intermediate powers of $QW_i$ are not
determined by the argument, even though finite, and so are
denoted by the dots in (\ref{eq: 3.12}). It will emerge that
the asymptotic properties of solitons that we seek do not
depend on these undetermined quantities. If $W_i$ tends
to zero, (\ref{eq: 3.12}) is dominated by the first term,
unity, and if $W_i$ tends to plus infinity, (\ref{eq: 3.12})
is dominated by the last term, given by the vertex
operator.

In particular, for the single soliton solution
(\ref{eq: 3.1}) in which $g(t)$ is given by a single
factor (\ref{eq: 3.3}), we see that the limits as $x$ tends to
$\pm \infty $ are respectively $ e^{-2\pi i
\lambda_i\cdot \lambda_j}$ and 1. This result assures that
the affine Toda field $\phi$ does interpolate degenerate
vacua at $x = \pm \infty$,  with the topological charge, $\Delta \phi$
satisfying
\beq
- {\beta \over 2 \pi i} \Delta \phi = \lambda_i + \Lambda_R (g)
 ,
\label{eq: 3.13}
\eeq
where we recall that $i$ labels the relevant soliton species.
Similar results apply to solutions describing any number of
solitons.

In this paper we shall address more refined questions
concerning the asymptotic behaviour of the N-soliton
solution (\ref{eq: 3.1}) and in particular determine the
lateral displacement of the soliton trajectories arising
from the collisions as described in section 2. We shall find
that the limited information outlined above is quite
sufficient for this  purpose, as the unknown coefficients
in  (\ref{eq: 3.12}) are irrelevant. What is important is the
number $X_{ik}(z_i,z_k)$ arising when the product of
two of the vertex operators (\ref{eq: 3.10}) is normal ordered \cite{OTUb,KOa}
\beq
Y^i_-(z_i) Y^i_+(z_i) Y^k_-(z_k) Y^k_+(z_k) =
   \left( X_{ik}(z_i,z_k) \right)^x  Y^i_-(z_i)Y^k_-(z_k)Y^i_+(z_i)Y^k_+(z_k)
\label{eq: 3.14}
\eeq
where $x$ is the level of the representation considered. It
was shown in \cite{OTUb} that
\beq
X_{ik}(z_i,z_k) = \prod_{p=1}^h{\left(z_i - e^{{2\pi i p
\over h}} z_k \right) ^{\gamma_i \cdot\sigma^p \gamma_k}}, \qquad
|z_i|>|z_k|,
\label{eq: 3.15}
\eeq
where the quantities $\sigma, \gamma_i , \gamma_k $ are
defined there.

By the commutation relations  (\ref{eq: 3.6}) we also find
\cite{KOa}
\bea
Y^i_+(z_i) \hat{F}^k(z_k) = X_{ik}(z_i,z_k)\hat{F}^k(z_k) Y^i_+(z_i),
   \label{eq: 3.16}  \\
\hat{F}^k(z_k) Y^i_-(z_i) = Y^i_-(z_i) \hat{F}^k(z_k) X_{ik}(z_i,z_k).
 \label{eq: 3.17}
\eea

We shall show in the next section that the quantity $X_{ik}$ appears
in the time delay  result (\ref{eq: 2.10}) and that when
(\ref{eq: 3.8}) is inserted it enjoys the properties required of
this physical interpretation (section 6).

\section{ Space-time trajectories of two colliding solitons}
\setcounter{equation}{0}
First we consider solutions with  two solitons and want to determine the
asymptotic form of
their trajectories in space-time and hence the lateral displacements defined in
section 2.
In the next section we consider collisions of more solitons, finding that the
two-soliton
result is the fundamental block, as expected in an integrable theory.
\par
The appropriate group element (\ref{eq: 3.2})  contains only two factors
\beq
g(t) \;= \; e^{ Q_2 W_{i(2)} (\th_2) \hat{F}^{i(2)}(\th_2) }
 e^{ Q_1 W_{i(1)} (\th_1) \hat{F}^{i(1)}(\th_1) }, \label{eq: 4.1}
\eeq
where $\th_1 > \th_2$ by (\ref{eq: 3.8}) and (\ref{eq: 3.15}). This inequality
means that soliton 1, of species $i(1)$,
moves faster than
soliton 2, of species $i(2)$. It must therefore start to the left of soliton 2
and eventually overtake
it, causing a collision whose outcome we wish to study.
\par
We shall do this by {\it ``tracking"}  each soliton in time. By tracking the
faster soliton 1 we mean that
we hold $W_{i(1)} (\th_1)$ fixed as time varies. As
\beq
W_{i(1)} (\th_1)  \; = \;  e^{\mu_{i(1)} \cosh \th_1 ( x - v_1 t ) },
\eeq
where $ v_1 = \tanh \th_1$ is the velocity of soliton 1, this means that, as t
varies, $x$ varies so as to hold $W_{i(1)}$
fixed, thereby remaining in the vicinity of the soliton, which is near
$x = v_1 t+x^0_1$. While $W_{i(1)}(\th_1)$ is
held fixed, that is $x - v_1t$ is fixed, the time dependence of
$W_{i(2)}(\th_2)$ is given by
\beq
W_{i(2)} (\th_2)  \; = \;  e^{\mu_{i(2)} \cosh \th_2 ( x - v_2 t ) } =
\hbox{const}  \;\;
e^{\mu_{i(2)} \cosh \th_2 ( v_1 - v_2  ) t }  \;\;  .
\eeq
Hence, in the past, $t \rightarrow - \infty$, $W_{i(2)}(\th_2)$ tends to 0 as $
v_1 > v_2$.
So, by (\ref{eq: 3.12})
\beq
e^{ Q_2 W_{i(2)} (\th_2) \hat{F}^{i(2)}(\th_2) }  \; \rightarrow \; 1\;\;  .
\eeq
Thus, by  (\ref{eq: 3.1}), in the past
\beq
e^{- \beta \lambda_j \cdot \phi } \; \rightarrow \;  \frac{  \langle \Lambda_j
| \;
e^{ Q_1 W_{i(1)} (\th_1) \hat{F}^{i(1)}(\th_1) }  \; | \Lambda_j \rangle }{
\langle \Lambda_0  | \;
e^{ Q_1 W_{i(1)} (\th_1) \hat{F}^{i(1)}(\th_1) } \;  | \Lambda_0 \rangle
^{m_{j}}  }
\label{eq: 4.5}
\eeq
which we recognise as a single soliton solution of species $i(1)$, velocity
$v_1$ and phase $\psi_1$. For
comparision, let us track soliton 1 in the two soliton solution into the
future,
so $t \rightarrow \infty$,
with $W_{i(1)} $ fixed, so that $W_{i(2)}$ tends to plus infinity. Now by
(\ref{eq: 3.12})  the
exponential is dominated by the highest non-vanishing power. In the numerator
of
 (\ref{eq: 3.10})
which has level $m_j$ this yields
\beq
e^{ Q_2 W_{i(2)} (\th_2) \hat{F}^{i(2)}(\th_2) }  \; \rightarrow \; e^{ - 2 \pi
i \lambda_{i(2)} \cdot \lambda_j}
( Q_2 W_{i(2)})^{m_j} Y_-^{i(2)} Y_+^{i(2)}  \;\;  .  \label{eq: 4.4}
\eeq
The factor $Y_-^{i(2)}$ annihilates to unity on the highest weight state
$\langle \Lambda_j |$
leaving the factor $Y_+^{i(2)}$, which would likewise annihilate to unity on
the right were it not for the
intervening factor  $e^{ Q_1 W_{i(1)} (\th_1) \hat{F}^{i(1)}(\th_1) }$.  By
(\ref{eq: 3.16})  these factors
can be interchanged if $Q_1$ is replaced by  $Q_1 X_{i(1) i(2) }(\th_{12})$.
Similar operations can be applied
to eliminate the vertex operator from the denominator. The large factors  $(
Q_2
W_{i(2)})^{m_j}$ cancel
between numerator and denominator, leaving in the future, $t \rightarrow
\infty$,
\beq
e^{- \beta \lambda_j \cdot \phi } \; \rightarrow \; e^{ - 2 \pi i
\lambda_{i(1)}
\cdot \lambda_j}
  \frac{  \langle \Lambda_j  | \;
e^{ Q_1 W_{i(1)} (\th_1) X_{i(1) i(2) }(\th_{12}) \hat{F}^{i(1)}(\th_1) }  \; |
\Lambda_j \rangle }{  \langle \Lambda_0  | \;
e^{ Q_1 W_{i(1)} (\th_1) X_{i(1) i(2) }(\th_{12}) \hat{F}^{i(1)}(\th_1) } \;  |
\Lambda_0 \rangle ^{m_{j}}  } \;\; .
\label{eq: 4.7}
\eeq
Again we recognise a single soliton solution of species $i(1)$, rapidity
$\th_1$ and phase $\psi_1$. The phase factor
preceding  (\ref{eq: 4.7}) is innocuous, representing a translation of $\phi$
by $ \frac{ 2 \pi i}{\beta} \lambda_{i(1)}$,
a symmetry of the theory. The other difference between (\ref{eq: 4.4}) and
(\ref{eq: 4.7}) is significant.
Since  $ X_{i(1) i(2) }(\th_{12})$ is real and positive, it means that
$Q_1$ has
acquired a factor $ X_{i(1) i(2) }(\th_{12})$
which changes its modulus (but not the phase), and hence $x_1^0$
(see (\ref{eq: 3.9}) ) in the
evolution from the past to the future.
The effect is that
\beq
\mu_{i(1)} \cosh \th_1 ( x - v_1 t ) \rightarrow \mu_{i(1)} \cosh \th_1 ( x -
v_1 t ) + \ln X_{i(1) i(2) }(\th_{12})
\eeq
so that the solution  (\ref{eq: 4.7})  differs from (\ref{eq: 4.4}) by a
translation in space-time. In particular
the trajectories  in space-time of the outgoing soliton is translated with
respect to the ingoing
soliton. Comparing with (\ref{eq: 2.3}) and the subsequent discussion we see
\beq
E_1 \Delta_{12} (x) \; = \; - \frac{ M_{i(1)}}{\mu_{i(1)} }  \;  \ln X_{i(1)
i(2) }(\th_{12})
\eeq
as the energy of the soliton is $E_1 = \mu_{i(1)} \cosh \th_1$. Employing the
mass formula (\ref{eq: 3.5}),
this equals
\beq
E_1 \Delta_{12} (x) \; = \; -  \;  \frac{2h}{ |  \beta^2 | }  \ln X_{i(1) i(2)
}(\th_{12})
\eeq
which is the announced result (\ref{eq: 2.10}) for the faster soliton.
\par
Now let us derive the corresponding result for the slower soliton 2,
by tracking it. As $W_{i(2)}$ is now
held fixed
\beq
W_{i(1)} (\th_1)  \; = \;   \hbox{const}  \;\;  e^{\mu_{i(1)} \cosh \th_1
( v_2 - v_1  ) t }  \;\;  .
\eeq
tends to $\infty$ and 0 in the past and future, respectively.
Thus, in the past,
\beq
e^{- \beta \lambda_j \cdot \phi } \; \rightarrow \; e^{ - 2 \pi i \lambda_{j}
\cdot \lambda_{i(1)}}
  \frac{  \langle \Lambda_j  | \;
e^{ Q_2 W_{i(2)} (\th_2) X_{i(1) i(2) }(\th_{12}) \hat{F}^{i(2)}(\th_2) }  \; |
\Lambda_j \rangle }{  \langle \Lambda_0  | \;
e^{ Q_2 W_{i(2)} (\th_2) X_{i(1) i(2) }(\th_{12}) \hat{F}^{i(2)}(\th_2) } \;  |
\Lambda_0 \rangle ^{m_{j}}  } \;\; .
\label{eq: 4.12}
\eeq
using (\ref{eq: 3.16}) with (\ref{eq: 3.17}), whereas in the future
\beq
e^{- \beta \lambda_j \cdot \phi } \; \rightarrow \;
  \frac{  \langle \Lambda_j  | \;
e^{ Q_2 W_{i(2)} (\th_2)  \hat{F}^{i(2)}(\th_2) }  \; | \Lambda_j \rangle }{
\langle \Lambda_0  | \;
e^{ Q_2 W_{i(2)} (\th_2)  \hat{F}^{i(2)}(\th_2) } \;  | \Lambda_0 \rangle
^{m_{j}}  } \;\; .
\label{eq: 4.13}
\eeq
So $Q_2  X_{i(1) i(2) }(\th_{12}) \rightarrow Q_2$  during the evolution and
\beq
E_2 \Delta_{21} (x) \; = \; + \;  \frac{2h}{ |  \beta^2 | }  \ln X_{i(1) i(2)
}(\th_{12}),
\eeq
thereby confirming (\ref{eq: 2.10}) for the slower soliton.

Notice that the solution considered describes only a transmission and not a
reflection of solitons. The only possible exception is when the species $i(1)$
and $i(2)$ coincide as do the phases $\psi_1$ and $\psi_2$. Then
we cannot tell whether the scattering is transmissive or reflective.

It is interesting to repeat the calculation with the order of the two
factors in (\ref{eq: 4.1}) reversed. The reader will find that the asymptotic
results
(\ref{eq: 4.5}), (\ref{eq: 4.7}), (\ref{eq: 4.12}) and (\ref{eq: 4.13}) are
unchanged, as is therefore the spatial displacement.

%*****************************************************************************
\section{ Space-time trajectories of any number of colliding solitons}
\setcounter{equation}{0}

It is not difficult to extend the preceding argument from the collision of two
solitons to the
collision of any number of various species. The interesting result is that the
total displacement
of the space-time trajectories of any chosen soliton is precisely the sum of
the
contributions
previously found for the collision of the chosen soliton with each of the
others.
This sum is
independent of the ordered sequence in which the chosen soliton collides with
the others. Hence this result
is the classical analogue of the Yang-Baxter \cite{YB} and bootstrap relations
\cite{ZZ} governing
quantum scattering in integrable theories.
\par
The solution  (\ref{eq: 3.1}) and (\ref{eq: 3.2}) describes N solitons. We
shall
choose to track one of these,
say soliton $m$, by remaining close to its trajectory in space-time as the past
evolves into the future. So
$ W_{i(m)}$,  (\ref{eq: 3.4}) , is held fixed and the other functions $
W_{i(n)}$ behave as
\beq
W_{i(n)} \; = \;     \hbox{const} \; e^{\mu_{i(n)} \cosh \th_2 ( v_m - v_n  ) t
} \;\; .
\eeq
Thus if soliton $n$ is slower than soliton $m$,  $W_{i(n)}$ tends to 0 in the
past
and $ \infty$ in the future. If
soliton $m$ is faster than soliton $n$, then the limits are reversed. Repeating
the arguments of the
preceding section we find that in the past
\beq
e^{- \beta \lambda_j \cdot \phi } \; \rightarrow \;  \exp\left( -2 \pi i
\sum\limits_{v_{k} < v_{m}}  \lambda_j \cdot
 \lambda_{i(k)} \right)   \frac{  \langle \Lambda_j  | \;
e^{ Q_m(I) W_{i(m)} (\th_m) \hat{F}^{i(m)}(\th_m) }  \; | \Lambda_j \rangle }{
\langle \Lambda_0  | \;
e^{ Q_m(I) W_{i(m)} (\th_m) \hat{F}^{i(m)}(\th_m) } \;  | \Lambda_0 \rangle
^{m_{j}}  } \label{eq: 5.2}
\eeq
where $Q_m(I) = Q_m \prod\limits_{v_{k} < v_{m}}  X_{i(m) i(k) } $.

In the future
\beq
e^{- \beta \lambda_j \cdot \phi } \; \rightarrow \;  \exp \left(-2 \pi i
\sum\limits_{v_{k} > v_{m}}  \lambda_j \cdot
 \lambda_{i(k)} \right)  \frac{  \langle \Lambda_j  | \;
e^{ Q_m(F) W_{i(m)} (\th_m) \hat{F}^{i(m)}(\th_m) }  \; | \Lambda_j \rangle }{
\langle \Lambda_0  | \;
e^{ Q_m(F) W_{i(m)} (\th_m) \hat{F}^{i(m)}(\th_m) } \;  | \Lambda_0 \rangle
^{m_{j}}  }
\eeq
where $Q_m(F) = Q_m \prod\limits_{v_{k} > v_{m}}  X_{i(m) i(k) } $.
This immediately yields the announced results:
\beq
E_m \Delta_m(x) = - p_m \Delta_m(t) = \frac{2h}{ |  \beta^2 | }
\left(   \sum\limits_{v_{k} > v_{m}}  \ln  X_{i(m) i(k) } -
\sum\limits_{v_{k} < v_{m}}  \ln  X_{i(m) i(k) }\right)
\label{eq: 5.4}
\eeq
where $  \Delta_m(x),  \Delta_m(t)$ denote the displacement  (\ref{eq: 2.3})
and
(\ref{eq: 2.5})  for the
$  m^ {\rm th}$ soliton.
\par
Notice that this result (\ref{eq: 5.4}) does not depend on the values of the
$Q_{i(n)}$, but only on the rapidities of the
solitons. Hence the temporal order of the scattering can be altered without
changing the rapidities of the
solitons.  Thus the overall displacement of the trajectories of soliton $m$,
(\ref{eq: 5.4}) is independent
of the order in which the collisions occurred.

The same procedure can be applied to each of the other $(N-1)$ solitons. In
this way we see how the $N$ soliton solution asymptotically contains the
$N$ single soliton solutions in both the past and future. The only alterations
in time are the lateral displacement of the trajectories specified by our
result (\ref{eq: 5.4}).

\section{Properties of the function $X_{ik}(\theta)$}
\setcounter{equation}{0}
The factor
\beq
X_{ik}(z_i,z_k)=\prod_{p=1}^h{\left(z_i - e^{{2\pi i p
\over h}} z_k \right) ^{\gamma_i\cdot\sigma^p \gamma_k}}
\label{eq: 6.1}
\eeq
arose \cite{OTUb,KOa} in the normal ordering of the product of the two vertex
operators, (\ref{eq: 3.14}),
$$
{(\hat{F}^i(z_i))^{m_j}\over m_{j}!} \quad {\rm and} \quad
{(\hat{F}^k(z_k))^{m_j}\over m_{j}!} $$

The exponents, $\gamma_i\cdot\sigma^p \gamma_k$, being scalar products of
roots of a simply laced Lie algebra, can only take the values
$0,\pm 1,\pm 2$. Thus $X_{ik}(z_i,z_k)$ can be analytically extended to a
meromorphic function of the
complex variables $z_i$ and $z_k$. $X_{ik}$ only  possesses a double pole if
$i=\bar{k}$, while the occurrence of simple poles is governed by Dorey's fusing
rule \cite{Dor1,OTUb,FLO}.

Using the first, (\ref{eq: 6.2a}), of the two facts
that
\beq
\sum_{p=1}^h \sigma^p\gamma_k=0  \label{eq: 6.2a} \eeq
and
\beq
\sum_{p=1}^h p\gamma_i\cdot\sigma^p \gamma_k\in h{Z\!\!\! Z},\label{eq: 6.2b}
\eeq
we can rewrite (\ref{eq: 6.1}) as
\beq
\prod_{p=1}^h (z_i z_k^{-1} - e^{2\pi i p \over h})^
{\gamma_i\cdot\sigma^p \gamma_k} \eeq
which means that $X_{ik}(z_i,z_k)$ depends on $z_i$ and $z_k$ only through
the ratio $z_i z_k^{-1}$. Furthermore using both  (\ref{eq: 6.2a})
and (\ref{eq: 6.2b}) we find that it exhibits the symmetry property
\beq X_{ik}(z_i,z_k)=X_{ki}(z_k,z_i), \eeq
which means that the vertex operators (\ref{eq: 3.10}) braid trivially.
This appears to be the explanation of our earlier observations that the order
of the soliton factors in (\ref{eq: 3.2}) is irrelevant.
Introducing the soliton rapidity $\theta_k$ via (\ref{eq: 3.8}),
$$ z_k=ie^{-\theta_k}e^{-{i\pi\over 2h}(1+c(k))} $$
we find that $X_{ik}$ can be expressed as a function of the rapidity
difference $\theta=\theta_i-\theta_k$:
\beq
X_{ik}(z_i,z_k)=\prod_{p=1}^h{\left(e^{-\theta} - e^{{\pi i
\over h}(2p+{c(i)-c(k)\over 2})}\right) ^{\gamma_i\cdot\sigma^p \gamma_k}}
= X_{ik}(\theta).
\label{eq: 6.5}
\eeq
Thus $X_{ik}(\th)$ is  Lorentz invariant since the relative rapidity is.
This is in accord with our result (\ref{eq: 2.10}).

The result (\ref{eq: 2.10}) which was established in the preceding
sections means that $X_{ik}(\th)$ has a space-time interpretation in terms
of the scattering of solitons. As a consequence it ought to be a real number
(when $\th$ is real) and exhibit some further symmetry properties.

We shall now check these properties explicitly, showing that $X_{ik}(\th)$
has period $2\pi i$,
\beq
X_{ik}(\th+2\pi i)=X_{ik}(\th), \label{eq: 6.7}\eeq
is symmetric in the sense
\beq
X_{ik}(\th)=X_{ki}(\th),\label{eq: 6.8} \eeq
is even in $\th$
\beq
X_{ik}(\th)=X_{ik}(-\th),\label{eq: 6.9} \eeq
takes values in the unit interval when $\th$ is real
\beq
0\leq X_{ik}(\th) < 1,\qquad \th\in { I \!\! R}, \label{eq: 6.10}\eeq
and obeys the ``crossing'' property
\beq
X_{\ib k}(\th)=(X_{ik}(\th+i\pi))^{-1} \label{eq: 6.11}\eeq
where $\ib$ denotes the anti-species of $i$.

Notice that the periodic property (\ref{eq: 6.7}) is already evident from
(\ref{eq: 6.5}). The symmetry property (\ref{eq: 6.8}) follows using the
identity
$$
\gamma_i\cdot\sigma^p \gamma_k = \gamma_k\cdot\sigma^{p'} \gamma_i,$$
where
$$ 2p+{c(i)-c(k)\over 2}=2p'+{c(k)-c(i)\over 2}. $$
To prove the evenness property (\ref{eq: 6.9}) note that
\begin{eqnarray*}
X_{ik}(-\th)&=&\prod_{p=1}^h{\left(e^{\theta} - e^{{\pi i
\over h}(2p+{c(i)-c(k)\over 2})}\right) ^{\gamma_i\cdot\sigma^p \gamma_k}} \\
&=&\prod_{p=1}^h{\left(e^{-\theta} - e^{{-\pi i
\over h}(2p+{c(i)-c(k)\over 2})}\right) ^{\gamma_i\cdot\sigma^p \gamma_k}}
\end{eqnarray*}
by (\ref{eq: 6.2a}) and (\ref{eq: 6.2b}). Now use $\gamma_i\cdot\sigma^p
\gamma_k=
\gamma_k\cdot\sigma^{-p}\gamma_i$ to recognize, on changing the dummy label
$p\rightarrow -p$, $X_{ki}(\th)$ which equals $X_{ik}(\th)$ by the symmetry
property (\ref{eq: 6.8}). To prove the reality property first note that
$$X_{ik}(\th^{*})^{*}=\prod_{p=1}^h{\left(e^{-\theta} - e^{{-\pi i
\over h}(2p+{c(i)-c(k)\over 2})}\right) ^{\gamma_i\cdot\sigma^p \gamma_k}},$$
which equals $X_{ik}(-\th)$ by (\ref{eq: 6.2a}) and (\ref{eq: 6.2b}) and hence
$X_{ik}(\th)$ by
evenness. Thus $X_{ik}(\th)$ is real when $\th$ is.

Now let us consider the possibility that $X_{ik}(\th)$ has zeroes or poles
when $\th$ is real. This is only possible when a factor vanishes, so that
both
$$\th=0, \qquad {\rm and}\quad p+{1\over 4}(c(i)-c(k))=0 \quad {\rm mod}\  h.$$
The second condition implies that $c(i)=c(k)$ and that $p=h$. When
$c(i)=c(k)$, $\gamma_i \cdot\sigma^{h}\gamma_k = \gamma_i \cdot\gamma_k$
vanishes unless $i=k$ when it equals $2$. So for real $\th $, $X_{ik}(\th)$
has no poles and the only zero occurs when $i=k$ and $\th=0$. We already knew
that $X_{ii}(0)$ had to be zero as it implies the nilpotency condition
$(\hat{F}^{i}(\th))^2=0$ at level $1$.

Now let us prove that $X_{ik}(\th)$ takes values in the unit interval. The
argument is intriguingly similar to that of section (4.5) of \cite{FO}
concerned with positivity properties of the affine Toda particle scattering
matrix.

Using the relation
$$ \gamma_i\cdot\sigma^p \gamma_k=\lambda_{i}\cdot\sigma^{-p+{c(k)-1\over 2}}
\gamma_k- \lambda_{i}\cdot\sigma^{-p+{c(k)+1\over 2}}\gamma_k,$$
we can rewrite (\ref{eq: 6.5}) as
\begin{eqnarray*}
X_{ik}(\th)& = &{\prod\limits_{p=1}^h \left(e^{-\theta} - e^{{\pi i
\over h}(2p+{c(i)-c(k)\over 2})}\right)
^{-\lambda_{i}\cdot\sigma^{-p+{c(k)+1\over 2
}}\gamma_k}\over
\prod\limits_{p=1}^h \left(e^{-\theta} - e^{{\pi i
\over h}(2p+{c(i)-c(k)\over 2})}\right)
^{-\lambda_{i}\cdot\sigma^{-p+{c(k)-1\over 2
}}\gamma_k} }  \\
& = &\prod_{p=1}^h \Biggl[{\sinh{1\over 2}(\th-{\pi i\over h}(2p-
{c(i)+c(k)\over 2} -1))\over
\sinh{1\over 2}(\th-{\pi i\over h}(2p-
{c(i)+c(k)\over 2} +1))} \Biggr]^{-\lambda_{i}\cdot\sigma^{p}\gamma_k}
\end{eqnarray*}
on relabelling the dummy index in order to gather the factors under a common
exponent. The factors can be further paired using the fact \cite{FO} that
$$
\lambda_i\cdot\sigma^p \gamma_k = -\lambda_i\cdot\sigma^{p'} \gamma_k,$$
where
$$p'=h+{c(i)+c(k)\over 2}-p.$$
Using this we can rewrite $X_{ik}(\th)$ as
\beq
X_{ik}(\th)=\prod_{p=a}^b \Biggl[{\cosh(\th)-\cos{\pi\over h}
(2p-{c(i)+c(k)\over 2} -1)\over
\cosh(\th)-\cos{\pi\over h}
(2p-{c(i)+c(k)\over 2} +1)}\Biggr]^{-\lambda_{i}\cdot\sigma^{p}\gamma_k}
\label{eq: 6.12}
\eeq
where $a={1+c(k)\over 2}$ and $b={h-1\over 2}+ {c(k)+c(\kb)\over 4}$. The
significance of the reduced range of $p$ is that, in it,
$$\lambda_{i}\cdot\sigma^{p}\gamma_k\leq 0.$$
Thus all the exponents in (\ref{eq: 6.12}) are positive. Thus in order to
prove (\ref{eq: 6.10}) it would be sufficient to show that each factor in
(\ref{eq: 6.12})
individually lies between $0$ and $1$. Because $\cosh\th\geq 1$ this is
ensured provided
\beq
1\geq \cos{\pi\over h}\biggl(2p-{c(i)+c(k)\over 2} -1\biggr) >
\cos{\pi\over h}\biggl(2p-{c(i)+c(k)\over 2} +1\biggr) > -1, \eeq
which follows from the fact that $\cos\phi$ is monotonically decreasing
from $1$ to $-1$ in the interval $0<\phi<\pi$, providing
$$0\leq 2p-{c(i)+c(k)\over 2} -1 < 2p-{c(i)+c(k)\over 2} +1 < h .$$
The smallest value of $2p-(c(i)+c(k))/2 -1$ occurs when
$p=a=(1+c(k))/2$ which is $(c(k)-c(i))/2$. This can only be negative when
$c(k)=-1=-c(i)$,
 so that $p=0$. In this case the exponent $\lambda_{i}\cdot\sigma^{p}\gamma_k=
\lambda_i\cdot\gamma_k =0$, as $i\neq k$, and the factor contributes unity.
A similar discussion applies to the upper limit $b$. Notice that expression
(\ref{eq: 6.12}) is explicitly real and even in $\th$.

Finally, using the relation \cite{FO},
$$\gamma_i=-\sigma^{-{h\over 2}-{c(i)-c(\ib)\over 4}}\gamma_{\ib},$$
we obtain the ``crossing'' property (\ref{eq: 6.11})
\begin{eqnarray*}
X_{ik}(\th+i\pi)&=&\prod_{p=1}^h \left(e^{-\theta} - e^{{\pi i
\over h}(2p-h+{c(i)-c(k)\over 2})}\right) ^{\gamma_i\cdot\sigma^p \gamma_k}\\
&=&\prod_{p=1}^h \left(e^{-\theta} - e^{{\pi i
\over h}(2p+{c(\ib)-c(i)\over 2})}\right)^{\gamma_{\ib}\cdot\sigma^p \gamma_k}
\\
&=&(X_{\ib k}(\th))^{-1}.
\end{eqnarray*}
Notice that, in agreement with the results of \cite{KOa}, the crossing property
involves the analytic continuation $\th\rightarrow\th + i\pi$ rather than
$\th\rightarrow i\pi-\th$
for the reasons explained by Coleman \cite{Cole2}, namely that the
semiclassical approximation breaks down on the imaginary rapidity axis.

It is further worth noting that by similar manipulations one can show that
$X_{ij}(\th)$ additionally satisfies the bootstrap equation \cite{Dor2,FO} and
thereby enhances the remarkable similarity in structure between $X_{ij}(\th)$
and the scattering matrix.

\section{Conclusions}

There are two main conclusions to our work and a number of comments. The
first result we have established is the intimate connection between the
space-time properties of the affine Toda solitons and the vertex operators
associated with them through the numerical function $X_{ik}(\theta)$ arising
when the product of the two vertex operators is normal ordered. This
connection is remarkable in view of the fact that these vertex operators
do not provide complete information concerning the solitons as they do not
completely determine the Kac-Moody group element (\ref{eq: 3.3}) but merely
yield (\ref{eq: 3.12}).

There is a well known result of Eisenbud and Wigner \cite{EW,W} relating the
time delay to the quantum mechanical scattering matrix in the semi-classical
approximation. The phase shift is obtained by integrating the time delay with
respect to energy, introducing a constant of integration proportional to the
number of bound states, presumably breathers in our context. This result
has been exploited in sine-Gordon theory \cite{JW,FK} but the breather spectrum
for general affine Toda field theory requires further study.

These results would presumably shed light on the intriguing similarity
in structure between $X_{ik}(\theta)$ and properties of particle scattering
matrix elements in affine Toda field theories \cite{Dor2,FO} as well as the
ideas of Corrigan and Dorey \cite{CD} for obtaining the S-matrices from the
braiding of vertex operators representing the Faddeev-Zamolodchikov operators
\cite{ZZ,Faddeev} .

Our second main result is that, since $X_{ik}(\th)$ takes values between $0$
and
$1$, it follows that the time delay experienced by any soliton in collision
with
any other in their centre of momentum frame is negative. This strongly suggests
that the forces between any two solitons is always attractive. This further
suggests that bound states (breathers) will form, though as far as we know,
this can only occur when the two solitons have equal mass and are
anti-species of each other. The only exception to the statement that forces
are attractive is when the two solitons involved are indistinguishable. Then
we can no longer recognize from the explicit solution that the scattering is
only a transmission as it could be reflective in this case. If the scattering
is considered to be reflective, the time advance can be ascribed to a repulsive
core. This is the accepted picture in sine-Gordon theory \cite{JW} where
independent arguments imply that the forces between indistinguishable
solitons are repulsive.

We should like to mention a delicate point here. We can distinguish the
outgoing solitons if they have different species or, if not, possess different
phases in $Q$. It is believed that the phase in $Q$, (\ref{eq: 3.9}), is
related to the topological quantum number of the soliton, (\ref{eq: 3.13}).
Unfortunately this connection has not been established in a satisfactorily
general way and the correspondence is not one to one. As mentioned above,
certain
discrete phases are forbidden in order that the soliton solutions be
nonsingular. The danger concerns possible zeros of the expectation values
of $g(t)$, (\ref{eq: 3.1}), or $\tau$-functions, as $x$ varies over space.
This leaves disconnected allowed ranges for the phase which seem to
correspond to specific values of the topological quantum number
(\ref{eq: 3.13}). The topological quantum number automatically takes
discrete values and is a continuous function of $Q$ except for discontinuities
occurring across the forbidden boundaries but the exact details are only
understood in the $su(n)$ case considered by McGhee \cite{McGhee}.
This remains an
outstanding issue. We should like to be able to say that two solitons of the
same species are distinguishable only if they carry different topological
quantum numbers and not just different phases, so two solitons could
be indistinguishable if they have the same topological quantum number but
different phases but this is not yet understood.

A second intriguing point concerns the repulsive core just mentioned for
indistinguishable solitons. This partly tallies with the fact that it is
impossible for two solitons of the same species to have the same rapidities.
 This is because
$$
\lim_{\theta_1,\theta_2\rightarrow\theta} e^{Q_{1}W_1\hat{F}^i(\th_1)}
e^{Q_{2}W_2\hat{F}^i(\th_2)}=e^{(Q_{1}+Q_{2})W\hat{F}^i(\th)} $$
which creates a single soliton rather than two. This phenomenon is familiar
in sine-Gordon theory where it is well known that, in the quantum theory, the
solitons are the fermions in the massive Thirring model
\cite{Skyrme,Cole,Mand}. It appears that
something like the exclusion principle is operating at the classical level.
The two results, repulsive core and exclusion principle, suggest that the
affine Toda solitons may also have a fermionic nature, but again much more
needs to be understood.

Finally we mention further remaining questions such as the extension to
non simply laced theories, and the question of time delays for the scattering
of breathers, once their spectrum is understood.
\vskip 0.2in
{\noindent\bf Acknowledgements}

\noindent We should like to thank Dr N. Dorey for discussions. AF wishes to
thank the Higher Education Funding Council of Wales for funding, PRJ thanks the
UK Science and Engineering Research Council and MACK thanks CNPq (Brazil).

\end{document}